\documentclass{iau}
\usepackage{graphicx}
\usepackage{amssymb}

\title[MOL-D a Database and a Web Service within SerVO and VAMDC] 
{Mol-D a Database and a Web Service within\\ the Serbian Virtual
    Observatory and the\\ Virtual Atomic and Molecular Data Centre}

\author[V.A.Sre{\' c}kovi{\' c}$^{1}$ et al. ]   
{Vladimir A. Sre{\' c}kovi{\' c}$^{1}$,
 Darko Jevremovi\'c$^{2}$, Veljko Vuj\v ci\'c$^{2,3}$, Ljubinko M. Ignjatovi{\' c}$^{1}$, Nenad Milovanovi{\' c}$^{2}$, Sanja Erkapi{\' c}$^{2}$ \& Milan S. Dimitrijevi{\' c}$^{2,4,5}$}

\affiliation{$^1$Institute of Physics Belgrade, University of Belgrade,\\  P.O. Box 57, 11001, Belgrade, Serbia \\ email: {\tt vlada@ipb.ac.rs} \\[\affilskip]
$^2$Astronomical Observatory, Volgina 7, 11060 Belgrade,
Serbia \\email: {\tt darko@aob.rs} \\[\affilskip]
$^3$Faculty of Organizational Sciences, Univesity of Belgrade, Serbia
\\[\affilskip]
$^4$Observatoire de Paris, 92195 Meudon Cedex, France \\email: {\tt mdimitrijevic@aob.rs} \\[\affilskip]
$^5$IHIS Techno experts, Batajni\v cki put 23, 11080 Zemun, Serbia}

\pubyear{2017}
\volume{325}  
\jname{Astroinformatics (AstroInfo16)}
\editors{A.C. Editor, B.D. Editor \& C.E. Editor, eds.}
\begin{document}

\maketitle

\begin{abstract}
In this contribution we report the current stage of the MOLecular Dissociation (MOL-D) database which is a web service
within the Serbian virtual observatory (SerVO) and node within Virtual Atomic and
Molecular Data Center (VAMDC).
MOL-D is an atomic and molecular (A\&M) database devoted to the modelling of stellar atmospheres,
laboratory plasmas, industrial plasmas etc.
The initial stage of development was done at the end of 2014, when
the service for data connected with
hydrogen and helium molecular ions was done.
In the next stage of the development of MOL-D, we include new
cross-sections and rate coefficients for processes which involve
species such as $X$H$^{+}$, where $X$ is atom of some metal. Data are important for
the exploring of the interstellar medium as well as for the
early Universe chemistry and for the modeling of stellar and solar
atmospheres. In this poster, we present our ongoing work and plans for
the future.
\end{abstract}

\firstsection 
\vspace*{-0.6 cm}
\section{Introduction}
In the era of large sky surveys and numerous large telescopes, data volumes have grown exponentially and increased from terabytes
into tens or hundreds of petabytes (\cite[Brunner et al. 2001]{bru01}, \cite[Djorgovski et al. 2013]{djo13}, \cite[York et al. 2000]{yor00})
and will increased even faster (\cite[LSST Science Collaboration et al. 2009]{lss09}).
At the same time, simultaneously a huge amount of information and data (produced by powerful supercomputer simulations)
are distributed through the global network infrastructure and stored in the networks of archives using new technologies (\cite[Allock et al. 2002]{all02}, \cite[Tate et al. 2016]{tat16}).
Consequently, the status of data-oriented science, research methods, algorithms, and techniques become very important.
Operative processing and scientific exploitation of such large data sets remains as one of the key motivations behind
the astronomical Virtual Observatory and a new emerging discipline Astroinformatics (\cite[Borne 2013]{bor13}, \cite[Brescia \& Longo 2013]{bre13}).
This has led to integration of computer science, physics, statistics, astrophysics etc.
In this contribution, we give an overview of the motivations, current stage, and technological principles of MOL-D database within Virtual Observatories.

\vspace*{-0.2 cm}
\section{Overview of MOL-D}
\label{sec:intro}
\begin{figure}[h!]
\begin{center}
\includegraphics[width=0.75\columnwidth,
height=0.48\columnwidth]{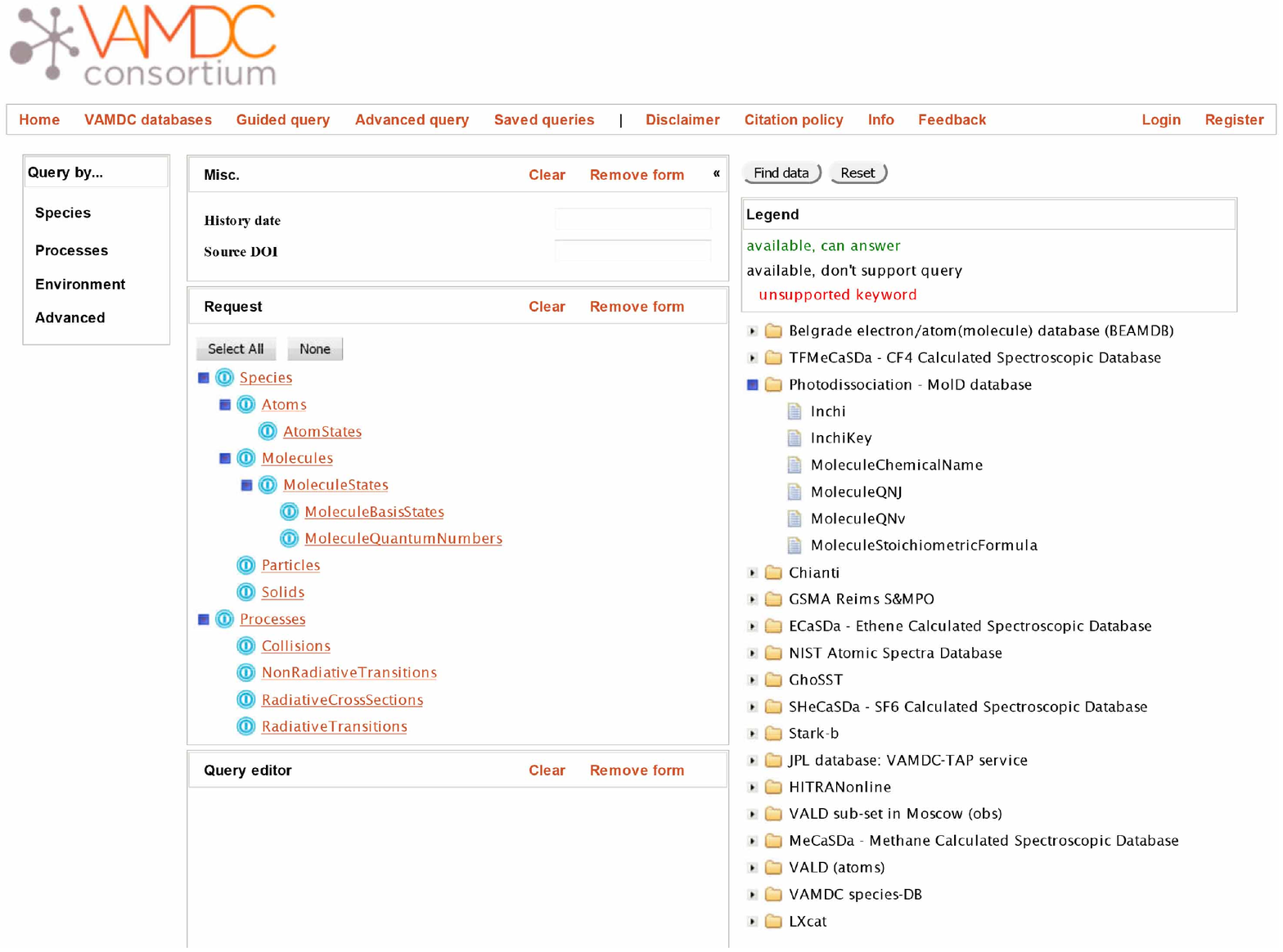} \caption{VAMDC (\cite[Dubernet et al. 2016]{dub16}) portal snapshot (http://www.portal.vamdc.eu).} \label{fig:snep2}
\end{center}
\end{figure}

Many fields in astronomy, atmospheric physics, chemistry, industry etc. depend on data for A\&M interaction processes.
A reliable exchange of such data has become crucial. In this context, we are developing the MOL-D database.
\begin{figure}[h!]
\begin{center}
\includegraphics[width=0.75\columnwidth,
height=0.51\columnwidth]{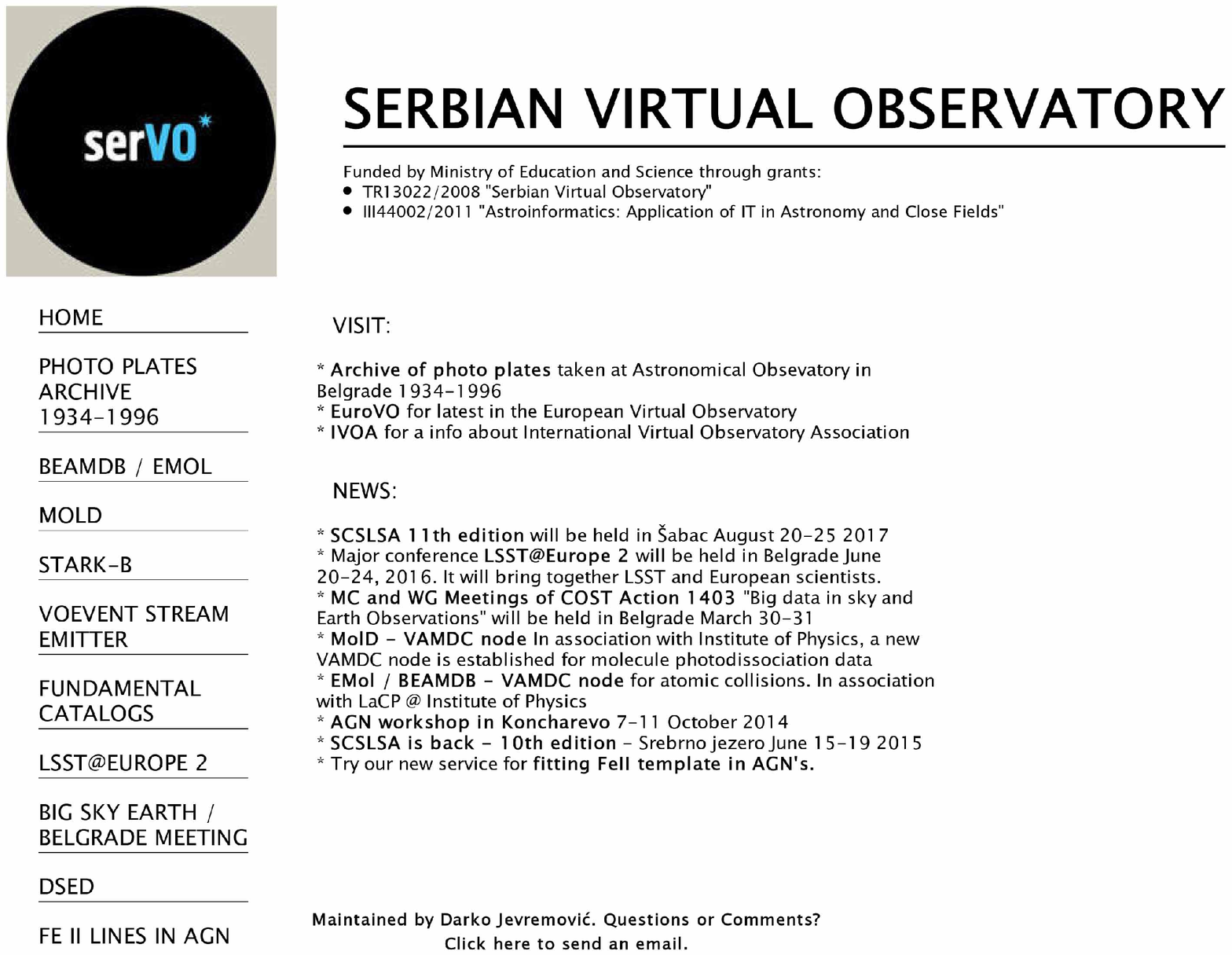} \caption{The home page of the SerVO (\cite[Jevremovi\' c et al. 2009]{jev09}).} \label{fig:snep1}
\end{center}
\end{figure}
MOL-D database is a collection of cross-sections and rate coefficients for specific collisional processes (\cite[Vuj\v ci\' c et al. 2015]{vuj15}).
It can be accessed via http://servo.aob.rs/mold/ of Serbian Virtual Observatory (SerVO, see Fig.\ref{fig:snep1})
or accessed as a web service http://portal.vamdc.eu/ which is part of the Virtual Atomic and Molecular Data Center
(VAMDC, see Fig.\ref{fig:snep2}).

\begin{figure}[ht]
\begin{center}
\includegraphics[width=0.81\columnwidth,
height=0.53\columnwidth]{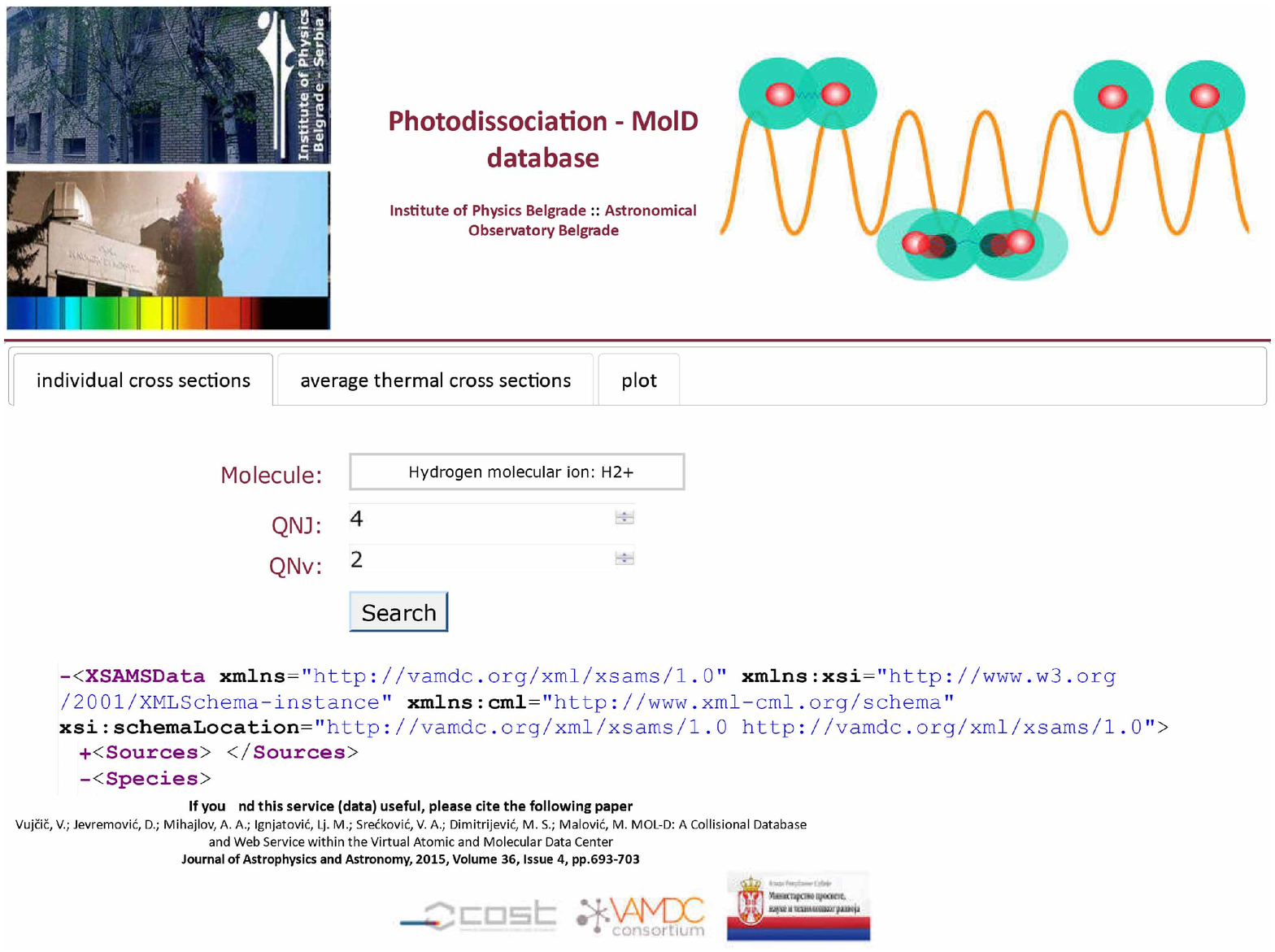} \caption{Homepage of MOL-D. Access to the data: the graphical interface. Data set is represented in XSAMS (Extensible Markup Language (XML) Schema for Atoms, Molecules and Solids) format.} \label{fig:snep3}
\end{center}
\end{figure}

The database contains photodissociation cross-sections for the individual ro-vibrational states of the diatomic molecular ions
and rate coefficients for the atom-Rydberg atom chemi-ionization and inverse electron-ion-atom chemi-recombination processes.
A graphical interface is provided at ’Homepage’ (Fig.\ref{fig:snep3}).
The web interface offers access to data for photodissociation (bound-free) cross-sections of
hydrogen H$_{2}^{+}$ and helium He$_{2}^{+}$ molecular ions (see e.g. figure \ref{fig:snep3}) as well as the corresponding
averaged thermal photodissociation cross-sections for the requested wavelength and temperature. All data sets are linked to the original article so that it is fully citable.
\vspace*{-0.4 cm}
\subsection{Level 2 Release}
At the end of 2016 MOL-D is in the \textit{stage 2} of development. We are including new cross-sections and rate coefficients
data for processes which involve species such as MgH$^{+}$, HeH$^{+}$, LiH$^{+}$, NaH$^{+}$, SiH$^{+}$ which are important for
exploring of the interstellar medium, the early Universe chemistry and for the
modeling of different stellar and solar atmospheres (\cite[Sre\' ckovi\' c et al. 2014]{sre14}, \cite[Coppola et al. 2016]{cop16}, \cite[Ignjatovi\' c et al. 2014a]{ign14a}, \cite[Ignjatovi\' c et al. 2014b]{ign14b}, \cite[Vuj\v ci\' c et al. 2015]{vuj15}).
Our plans are incremental inclusion of data from our papers concerning atomic and molecular processes
important for modeling different stellar atmospheres and laboratory plasmas as they become published.
Along with database updates, we intend to develop new web utilities and interfaces for SerVO MOL-D website.

\vspace*{-0.5 cm}
\section{Database technical description }
VAMDC is an international consortium which has built a well documented, secure, flexible interoperable
e-science platform permitting an automated exchange of atomic and molecular data.
VAMDC e-infrastructure defines protocols for retrieving remote data as well as format for representing these data.
The ultimate goal is interoperability of  the data along various distributed nodes.
Access to the MOL-D data is possible via Table Access Protocol (TAP), a Virtual Observatory standard of a web service
or via AJAX (Asynchronous JavaScript and XML)-enabled web interface (http://servo.aob.rs/mold).
Both queries return data in XSAMS (XML Schema for Atoms, Molecules and Solids) format (\cite[Braams et al. 2016]{bra16}).
The XSAMS schema provides a framework for a structured presentation of atomic, molecular,
and particle-solid-interaction data in an XML file.
Underlying application is written in Django, a Python web framework and represents a customization and
extension of VAMDC‘s NodeSoftware. Additional on-site utilities include:
data selection based on molecule name and QNJ/QNv numbers;
average thermal cross section calculation based on temperature for a specific molecule and wavelength;
plotting average thermal cross sections along available wavelengths for a given temperature.

\vspace*{-0.5 cm}
\section{Concluding remarks}
\label{sec:RD}

Atomic and molecular databases in today's science have become essential for
modeling and interpretation of data provided by observations and laboratory measurements
(\cite[Sre\' ckovi\' c et al. 2014]{sre14}, \cite[Mihajlov et al. 2012]{mih12}, \cite[Mihajlov et al. 2013]{mih13}, \cite[Marinkovi\' c et al. 2015]{mar15}).
Nowadays, very important resources of such data are dissipated, presented in different,
non-standardized ways, available through a variety of  highly  specialized  and  often
not very well documented interfaces. As a consequence the full exploitation and interconnection
of all available data is limited and the free exchange of these data requires the definition of some
standards and tools that help users and producers in this process.
The continuation of such developments and services is crucial.
This is the purpose of further development of MOL-D database within SerVO and VAMDC.

\vspace*{-0.2 cm}
\begin{acknowledgments}
This work was supported by the Ministry of Education, Science and Technological Development of the
Republic of Serbia Grants III44002 (Astroinformatics: Application of IT in Astronomy and Close Fields)
and OI176002 (Influence of collisional processes on the astrophysical plasma spectra).
A part of this work has been supported by VAMDC.
\end{acknowledgments}

\end{document}